\documentclass[twocolumn,superscriptaddress,showpacs,aps,amsmath,amssymb,pra]{revtex4}

\usepackage{etex}
\usepackage{bbm}
\usepackage{epsfig}
\usepackage[utf8]{inputenc}

\usepackage{bm}
\usepackage{graphicx,color}

\newcommand{\be}{\begin{equation}}
\newcommand{\ee}{\end{equation}}
\newcommand{\bea}{\begin{eqnarray}}
\newcommand{\eea}{\end{eqnarray}}
\newcommand{\bsube}{\begin{subequations}}
\newcommand{\esube}{\end{subequations}}

\newcommand{\Havg}{\overline{H}}
\newcommand{\id}{\mathbbm{1}}
\newcommand{\braket}[2]{{\left\langle{#1}\vert{#2}\right\rangle}}

\input{Qcircuit}

\begin{document}

\title{Pulse-controlled quantum gate sequences on a strongly coupled qubit chain}

\author{Holger Frydrych}
\email{holger.frydrych@physik.tu-darmstadt.de}
\affiliation{Institut für Angewandte Physik, Technische Universität Darmstadt, D-64289 Darmstadt}

\author{Michael Marthaler}
\affiliation {Institut f\"ur Theoretische Festk\"orperphysik,
      Karlsruhe Institute of Technology (KIT), 76131 Karlsruhe, Germany}
      
\author{Gernot Alber}
\affiliation{Institut für Angewandte Physik, Technische Universität Darmstadt, D-64289 Darmstadt}


\date{\today}

\begin{abstract}
We propose a selective dynamical decoupling scheme on a chain of permanently coupled qubits with XX type interactions, which is capable of dynamically suppressing any coupling in the chain by applying sequences of local pulses to the individual qubits. We demonstrate that high-fidelity single- and two-qubit gates can be achieved by this procedure and that sequences of gates can be implemented by this pulse control alone. We discuss the applicability and physical limitations of our model specifically for 
strongly coupled superconducting flux qubits. Since dynamically modifying the couplings between flux qubits is challenging, they are a natural candidate for our approach.
\end{abstract}


\maketitle

\section{Introduction}
\label{sec:intro}

Current implementations of qubits are typically either well isolated from noise, but difficult to couple, or strongly coupled, but difficult to isolate.
To achieve the goal of building a working quantum computer for the tasks of quantum simulation and, eventually, quantum computation, 
we require an architecture that is capable of strongly coupling qubits to implement fast multi-qubit gates, but that can also isolate qubits from each other and the environment when no gate operation is performed. 
The trade-off between strong coupling and isolation should be optimized to maximize the ratio between decoherence time and gate operation time.

In quantum optics, extensive work has been done using trapped ions or atoms as qubits, and scalable architectures that can trap and address a large number of qubits simultaneously exist \cite{Schlosser_Birkl_Atoms_in_Microlens_Array}. These qubits feature excellent coherence times, yet the implementation of two-qubit gates in these architectures is still a topic of ongoing research, although recently promising proposals were made in this regard \cite{Anderlini_Exchange_Interaction, Wilk_Rydberg_Blockade, Isenhower_CNOT_Neutral_Atoms}.

Likewise, we have seen significant progress in solid state qubit architectures  \cite{Kane_Computer_Single_Qubit_great_coherence,NV_Center_Wrachtrup,Nori_Review}, 
and there exist promising candidates for scalable systems. Gate-defined spin qubits \cite{Exponential_Increase_of_Coherence_Devoret,Review_Spin_qubits} feature excellent coherence properties \cite{Spin_Qubit_good_coherene_Bluhm}, but coupling two qubits remains a challenge despite proposals for efficient coupling \cite{cQED_Spin_Qubits_Jin,Experimental_Coupling_of_Spin_Qubits_Yacoby}.
For superconducting qubits, both indirect coupling via a resonator \cite{cQED_three_qubits_coupled_Schoelkopf} and direct capacitive 
coupling of detuned qubits  \cite{capacative_coupling_superconuctingqubits_Martinis} have been demonstrated and are comparatively easy to realize. While in principle superconducting
qubits can be coupled very strongly, the small anharmonicity of some designs limits the possible coupling strength \cite{Multilevel_spectroscopy_Transmon}.
Very recently dynamically protected superconducting qubits have been proposed \cite{Devoret_Proposal_Protected}
and demonstrated \cite{Devoret_Exp_Protected},
which rely on a parametrically driven cavity \cite{Devoret_Parametric}
in the quantum regime \cite{Devoret_Siddiqi_Quantum_Review,Driven_Quantum_Spectrum,Driven_Quantum_Tunneling}.
In this design the qubits are well protected, but direct coupling of two qubits might remain a challenge.

Good coherence properties were achieved for flux qubits \cite{Good_Flux_Qubit_Catalani}, a particular type of superconducting qubit with a very large anharmonicity. 
This anharmonicity allows them to be strongly coupled \cite{Coupling_two_flux_qubits}, 
which makes them particularly interesting for the implementation of fast two-qubit gates.  Larger circuits containing many weakly coupled flux qubits have 
already been demonstrated \cite{Metamaterial}.
However, their tunability is limited by the need for an optimal operating point, which makes it difficult to isolate the qubits when no gate operation should be performed.

In this article, we attempt to overcome the isolation problem of flux qubits and other similarly strongly coupled qubit systems by an alternative ansatz. We study a qubit chain with always present nearest-neighbor couplings and make use of a pulse generator to exert external control on the qubits with the aim of suppressing unwanted qubit couplings.  We demonstrate in numerical simulations that this simple pulse control enables us to implement a sequence of entangling gate operations on the qubit chain to entangle all the qubits in the chain in a GHZ state \cite{ghz89} with high fidelity. We thus show that a system of strongly coupled flux qubits may be used for universal quantum computation purposes without the need to control the qubit couplings.

Our pulse control is based on dynamical decoupling \cite{vkl99}, which is a generalization of techniques developed in the nuclear magnetic resonance (NMR) community \cite{hahn50, Carr, Meiboom, Haeberlen}. It makes use of external control pulses being applied in
rapid succession to the system in question. With a carefully designed control sequence it is possible to eliminate 
(parts of) a Hamiltonian interaction up to a certain order. 
Dynamical decoupling has been successfully implemented in numerous experiments to protect qubit states from the effects of decoherence \cite{mtabplb06,fsl05,budsib09,lgdvt11,hhh13}. 
For our purposes, we are interested in selectively decoupling only certain interactions between qubits while keeping others alive,
a possibility proposed already by Viola {\it et al.} \cite{vlk99}. 
A particularly simple to handle subset of decoupling schemes applicable to networks of qubits employs only Pauli pulses to individual qubits. 
Several different construction methods for such Pauli operator schemes exist \cite{sm01,leu02,rw06,wrjb02b,fab14}.
In dynamical decoupling, it is typically assumed as a first approximation that the applied control pulses are instantaneous and unitary. In our numerical calculations, we go beyond this approximation by simulating realistic pulse lengths. To deal with such bounded controls, advanced decoupling techniques in the form of Eulerian decoupling \cite{Viola_Euler_Decoupling} and dynamically corrected gates \cite{khodjasteh_prl,khodjasteh_dcg} have been developed, which we will make use of.

The paper is organized as follows: In section \ref{sec:model} we present the physical model of
our qubit chain and the type of control we have over the system. 
Section \ref{sec:iswap_gate} explains how a two-qubit iSWAP gate can be implemented with the help of dynamical decoupling. Decoupling basics are explained and a decoupling sequence for this particular task is developed. Numerical results for the achievable fidelity are presented. In section \ref{sec:sq_gates} we then 
look at how to implement single-qubit gates with high fidelity, where numerical simulations were conducted to verify the achievable fidelities. Finally, in section
\ref{sec:cnsgate} we introduce the CNS gate and use this gate to entangle all the qubits in our chain in a GHZ state. We calculate numerical results for the achievable GHZ state fidelity for different numbers $N$ of qubits and also look at how much of an impact disorder has on the fidelity.

\section{The coupled qubit system model} \label{sec:model}
We consider a system of $N$ qubits in a chain with nearest-neighbor couplings described by the Hamiltonian 
\begin{equation}
H_0 = \frac{1}{2} \sum_{i=1}^N \epsilon_i \sigma_3^{(i)} - g \sum_{i=1}^{N-1} \sigma_1^{(i)} \sigma_1^{(i+1)},  \label{eq:hamiltonian_0}
\end{equation}
where the $\sigma_a^{(i)}$ are the Pauli operators applied to the $i$-th qubit, and $\epsilon_i$ are the qubits' eigenenergies. The coupling between the
qubits is assumed to be uniform and characterized by the coupling strength $g$. 
This model is strongly inspired by a system of coupled flux qubits \cite{Coupling_two_flux_qubits}, however, alternative qubit designs exist which are also described by this Hamiltonian. Additionally, in our model there is a pulse generator with frequency $\omega$ which can exert external control on the qubits, and in the case of flux qubits is implemented as a microwave emitter. It is described by the control Hamiltonian
\begin{equation}
H_{c0}(t) =  \sum_{i=1}^N f_i(t) \sigma_1^{(i)} \cos(\omega t+\varphi_i(t))
\end{equation}
and is governed by the pulse amplitudes $f_i(t)$ and phases $\varphi_i(t)$, which can be controlled for each qubit individually.

It is convenient to switch to a rotating frame by transforming to the interaction picture given by the unitary operator $U_\omega(t)=\exp(i\omega\sum_i\sigma_3^{(i)} t/2 )$. In the rotating frame and with the rotating wave approximation, the system and control Hamiltonians equal
\begin{align}
H =& \frac{1}{2} \sum_{i=1}^N \Delta_i \sigma_3^{(i)} - \frac{g}{2} \sum_{i=1}^{N-1} \left( \sigma_1^{(i)} \sigma_1^{(i+1)} + \sigma_2^{(i)} \sigma_2^{(i+1)} \right),  \notag \\
H_c(t) =& \frac{1}{2} \sum_{i=1}^N f_i(t) \left(\cos(\varphi_i(t)) \sigma_1^{(i)} + \sin(\varphi_i(t)) \sigma_2^{(i)} \right). \label{eq:hamiltonian}
\end{align}
The $\Delta_i=\epsilon_i-\omega$ indicate the detuning between the individual qubits' 
eigenenergies and the frequency of the driving field and should ideally be zero for our purposes. If the eigenenergies are different, then we have disorder, which can disrupt the gate operations we intend to implement in the following. However, as we will see, our approach is robust to disorder due to our use of decoupling, as long as the $\Delta_i$ do not become too large.

\section{Implementing the two-qubit $\text{iSWAP}$ gate by selective decoupling} \label{sec:iswap_gate}
The coupling between the qubits according to \eqref{eq:hamiltonian} is of $XX$ type. Schuch and Siewert \cite{ss03} studied natural gate operations resulting from such an interaction. They showed that, after an interaction time $T = \pi/(2g)$, this type of coupling between two qubits produces a unitary $\text{iSWAP}$ gate:
\begin{align}
U_\text{iSWAP} :=& \exp\left[ iT \frac{g}{2} \left( \sigma_1^{(i)} \sigma_1^{(i+1)} + \sigma_2^{(i)} \sigma_2^{(i+1)} \right) \right] .
\end{align}
This gate, like the better known SWAP gate, exchanges the state of two qubits, but introduces an additional phase on the swapped qubit states. However, in our model we have additional couplings to the qubits $(i-1)$ and $(i+2)$ as well as the disorder terms $\Delta_i \sigma_3^{(i)}$ and $\Delta_{i+1} \sigma_3^{(i+1)}$. In order to succesfully use the natural couplings to implement the iSWAP gate, we need to isolate the two qubits involved in the gate operation.
Traditionally, we would thus require switching off any interactions which are not currently needed, but this process is complicated and often limits the achievable interaction strength 
$g$. Instead, we will employ dynamical decoupling to suppress the effects of individual couplings as needed.

\subsection{Dynamical decoupling basics} \label{sec:dd_basics}
In dynamical decoupling, the natural evolution of the $N$-qubit chain under the acting Hamiltonian $H$ is modified in a controlled fashion by the external control Hamiltonian $H_c(t)$. In our case, the pulse generator will be activated periodically at times $t_j$ for a short time $t_p$ to implement a sequence of pulses
\begin{equation}
p_j = U_c(t_j, t_j+t_p) = \mathcal{T} \exp\left(-i \int_{t_j}^{t_j+t_p} dt H_c(t) \right),
\end{equation}
where $\mathcal{T}$ denotes the Dyson time-ordering operator and $p_j$ is a unitary operator representing the $j$-th pulse of the sequence. However, the implementation of the pulse is disturbed by the acting Hamiltonian $H$, so that we get an imperfect pulse of the form
\begin{align}
\tilde{p}_j &= \mathcal{T} \exp\left(-i \int_{t_j}^{t_j+t_p} dt (H_c(t) + H) \right) = p_j p_j^\dag \tilde{p}_j \notag \\
&= p_j \mathcal{T} \exp\left(-i \int_{t_j}^{t_j+t_p} dt U_c^\dag(t_j,t) H U_c(t_j,t) \right) \notag \\
&\equiv p_j e^{-i \Phi_j}.
\end{align}
If we assume that after each decoupling pulse there is a time $\tau$ of free evolution under the Hamiltonian $H$, then by introducing the unitary operators $g_j = p_j p_{j-1} \cdots p_0$, the resulting time evolution $U(t)$ after $M$ pulses can be written as
\begin{align} \label{eq:dec_evolution}
U(M \tau) =& \tilde{p}_M e^{-iH\tau} \tilde{p}_{M-1} \cdots \tilde{p}_1 e^{-iH\tau}  \notag \\
=& g_M \left(g_{M-1}^\dag e^{-i\Phi_M} e^{-iH\tau} g_{M-1} \right) \notag \\
& \cdots \left(g_{0}^\dag e^{-i\Phi_1} e^{-iH\tau} g_{0} \right)  \notag \\
=& g_M e^{-i g_{M-1}^\dag \Phi_M g_{M-1}} e^{-i g_{M-1}^\dag H g_{M-1} \tau} \notag \\
& \cdots e^{-ig_0^\dag \Phi_1 g_0} e^{-i g_0^\dag H g_0 \tau}.
\end{align}
It is customary to enforce the cyclic condition $g_M = g_0 = \id$ by an appropriate choice of the decoupling pulses $p_j$. We can now define an average Hamiltonian $\Havg$ which leads to the same time evolution after the time $M\tau$, i.e.,
\begin{equation}
U(M \tau) \equiv e^{-i \Havg M \tau}.
\end{equation}
By performing a Magnus expansion \cite{magnus54}, the average Hamiltonian $\Havg$ is expanded in powers of the pulse distance $\tau$, i.e.,
\begin{equation}
\Havg = \Havg^{[0]} + \Havg^{[1]} + \Havg^{[2]} + \dots ,
\end{equation}
where the lowest order is found to be
\begin{equation} \label{eq:decoupling_condition}
\Havg^{[0]} = \frac{1}{M} \sum_{j=0}^{M-1} g_j^\dag \left(H + \frac{1}{\tau} \Phi_{j+1}^{[0]}\right) g_j .
\end{equation}
Here, $\Phi_j^{[0]}$ is the lowest order of the Magnus expansion of the error operator $\Phi_j$, which is given by
\begin{equation}
\Phi_j^{[0]} = \int_0^{t_{p_j}} dt U_{p_j}^\dag(t) H U_{p_j}(t) .
\end{equation}

Our goal is to selectively remove couplings between specific qubit pairs in the lowest order $\Havg^{[0]}$ of the average Hamiltonian and to keep all others, while simultaneously suppressing the effects of the disorder terms $\Delta_i \sigma_3^{(i)}$ and the pulse errors $\Phi_j^{[0]}$. We call a set of $M$ operators $\{g_j\}_{j=0}^{M-1}$ 
a decoupling scheme if it fulfils this purpose. Note that the higher orders of $\Havg$ are typically non-zero and remain as errors.

\subsection{Decoupling an individual qubit} \label{sec:single_qubit}
As a first step, we will discuss decoupling of a single qubit on the chain with the goal of freezing the evolution of that qubit's state. For a single qubit, there exists a particular decoupling scheme 
\begin{equation} \label{eq:annihilator}
\left\{ g_0 = \id, \; g_1 = \sigma_1, \; g_2 = \sigma_3, \; g_3 = \sigma_2 \right\},
\end{equation}
which has the property that for any traceless Hermitian operator $P$ acting on the subspace of the qubit,
\begin{equation}
\sum_{i=0}^3 g_i^\dag P g_i = 0.
\end{equation}
It can be implemented solely with the help of $\sigma_1$ and $\sigma_2$ pulses to the qubit (meaning $\pi$ pulses around the $X$ or $Y$ axis, respectively), by the sequence
\begin{align}
g_0 =& \id \stackrel{p_1=\sigma_1}{\longrightarrow} \sigma_1 \stackrel{\sigma_2}{\longrightarrow} \sigma_3 \stackrel{\sigma_1}{\longrightarrow} \sigma_2 \stackrel{\sigma_2}{\longrightarrow} \id .
\end{align}
If we insert this decoupling scheme into \eqref{eq:decoupling_condition}, it will eliminate all parts of the Hamiltonian $H$ acting on the qubit in the lowest order, effectively decoupling the qubit from the rest of the chain. Unfortunately, the pulse errors $\Phi_j$ depend on the particular pulses $p_j$ and are thus not eliminated by this decoupling scheme.

\begin{figure}
\includegraphics[width=7cm]{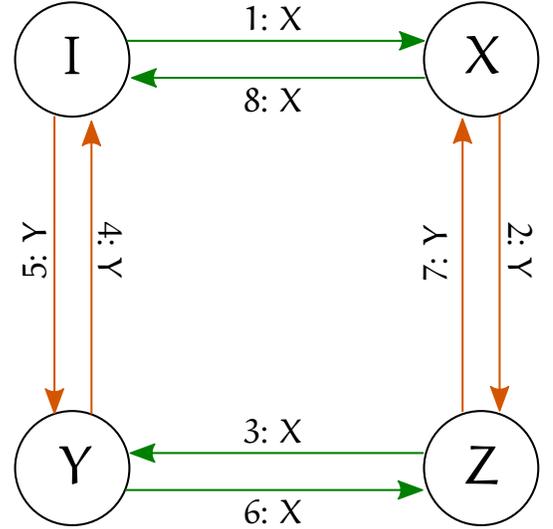}
\caption{\label{fig:euler} A Eulerian path decoupling sequence for a single qubit. $I,X,Y,Z$ correspond to the Pauli operators $\id$, $\sigma_1$, $\sigma_2$ and $\sigma_3$. The vertices represent the decoupling operators $g_j$, the directed edges denote the transitions between the $g_j$ due to the decoupling pulses $p_j$.}
\end{figure}

There is a trick to construct a decoupling sequence from the scheme in Eq.  \eqref{eq:annihilator} which not only eliminates $H$, but also the pulse errors $\Phi_j$ in \eqref{eq:decoupling_condition}. This method is called Eulerian path decoupling \cite{Viola_Euler_Decoupling} and proceeds as follows. A graph is constructed from the scheme \eqref{eq:annihilator} where the scheme operators $g_j$ are taken as the vertices of that graph and a directed edge is placed between two operators $g_a$, $g_b$ if $\sigma_1 g_a = g_b$ or $\sigma_2 g_a = g_b$, up to a phase factor. A Eulerian path through this graph is a path which visits every edge of the graph exactly once. A particular Eulerian path is depicted in figure \ref{fig:euler}, which results in the following decoupling sequence:
\begin{align} \label{eq:euler_seq}
g_0 =& \id \stackrel{p_1=\sigma_1}{\longrightarrow} \sigma_1 \stackrel{\sigma_2}{\longrightarrow} \sigma_3 \stackrel{\sigma_1}{\longrightarrow} \sigma_2 \stackrel{\sigma_2}{\longrightarrow} \notag \\
& \id \stackrel{\sigma_2}{\longrightarrow} \sigma_2 \stackrel{\sigma_1}{\longrightarrow} \sigma_3 \stackrel{\sigma_2}{\longrightarrow} \sigma_1 \stackrel{\sigma_1}{\longrightarrow} \id.
\end{align}
It corresponds to two consecutive applications of the original scheme \eqref{eq:annihilator}, but with different orders of the scheme operators $g_j$. If inserted into \eqref{eq:decoupling_condition}, we get
\begin{equation}
\Havg^{[0]} = \frac{1}{8} \sum_{\substack{g_j \in \\ \{\id, \sigma_1,\sigma_2,\sigma_3\}}} g_j^\dag \left(2H + \frac{1}{\tau}(\Phi_1^{[0]} + \Phi_2^{[0]})\right) g_j = 0.
\end{equation}
The remaining orders of the average Hamiltonian $\Havg$ are of order $\mathcal{O}(||H||^2 \tau) + \mathcal{O}(||\Phi_j||^2)$.

\subsection{Selective decoupling on the qubit chain} \label{sec:combined}
The decoupling sequence discussed in section \ref{sec:single_qubit} can isolate a single qubit from the chain. We need to extend this sequence to the whole chain in such a way that we can selectively decouple only certain qubit couplings while keeping others alive. In \cite{fab14} we found that if we extend the original decoupling scheme \eqref{eq:annihilator} to two qubits in the following way,
\begin{align} \label{eq:tq_keep}
g_0 &= \id \otimes \id, & g_1 &= \sigma_1 \otimes \sigma_1, \notag \\ 
g_2 &= \sigma_2 \otimes \sigma_2, & g_3 &= \sigma_3 \otimes \sigma_3 ,
\end{align}
then in \eqref{eq:decoupling_condition} it will keep the Heisenberg-type coupling terms between these two qubits intact while still eliminating the disorder terms $\Delta_i \sigma_3^{(i)}$. On the other hand, if instead we choose
\begin{align} \label{eq:tq_eliminate}
g_0 &= \id \otimes \id, & g_1 &= \sigma_1 \otimes \sigma_2, \notag \\ 
g_2 &= \sigma_2 \otimes \sigma_1, & g_3 &= \sigma_3 \otimes \sigma_3 ,
\end{align}
then the Heisenberg-type couplings in \eqref{eq:decoupling_condition} are eliminated between these two qubits. The first scheme can be realised by applying the pulse sequence $XYXY$ on both qubits simultaneously, while the latter one applies $XYXY$ to one qubit and $YXYX$ to the other. 

It is straight-forward to extend these schemes to the whole qubit chain. On each qubit, we apply alternating $XY$ pulses. Neighbouring qubit pairs whose interaction should be kept alive will employ $XY$ pulses in the same order, whereas between qubit pairs whose interaction should be decoupled, we use alternating pulse sequences. 
For example, if we wanted to protect the interaction between the first and last qubit pairs on a 5-qubit chain, but eliminate the couplings with the middle qubit, we would use the following pulse sequence:
\begin{align} \label{eq:chain_combined}
p_1 &= \sigma_1^{(1)} \sigma_1^{(2)} \sigma_2^{(3)} \sigma_1^{(4)} \sigma_1^{(5)} \notag \\
p_2 &= \sigma_2^{(1)} \sigma_2^{(2)} \sigma_1^{(3)} \sigma_2^{(4)} \sigma_2^{(5)} \notag \\
p_3 &= p_0  \notag \\
p_4 &= p_1 .
\end{align}

While the previously explained extension allows us to selectively decouple certain qubit couplings from the Hamiltonian $H$ in \eqref{eq:decoupling_condition}, it does not eliminate the pulse errors $\Phi_j^{[0]}$. A simple modification to \eqref{eq:chain_combined} sees us adding additional pulses
\begin{equation}
p_5 = p_4, \quad p_6 = p_3, \quad p_7 = p_2, \quad p_8 = p_1,
\end{equation}
with the effect that on each individual qubit we now have a Eulerian path decoupling sequence as in \eqref{eq:euler_seq}, without changing the effects of the sequence on $H$ in the lowest order \eqref{eq:decoupling_condition}. This is the final decoupling sequence which we will use to implement our quantum gates, and we will see in numerical simulations that it produces sufficiently high fidelities. However, we should point out that, due to the extension of the sequence to the whole chain, the pulse errors are not fully eliminated in the lowest order even with the Eulerian path modification. The reason is that the original schemes \eqref{eq:tq_keep} and \eqref{eq:tq_eliminate} only eliminate certain Hermitian operators on the two-qubit subspace, but not all of them. A more sophisticated approach is outlined in \cite{khodjasteh_dcg} which eliminates errors completely (in the lowest order), however, it requires 64 pulses instead of 8 and thus has a significantly longer implementation time.

\subsection{The $\text{iSWAP}$ gate and physical limitations} \label{sec:iswap}
\begin{figure}
\begin{center}
\includegraphics[width=8cm]{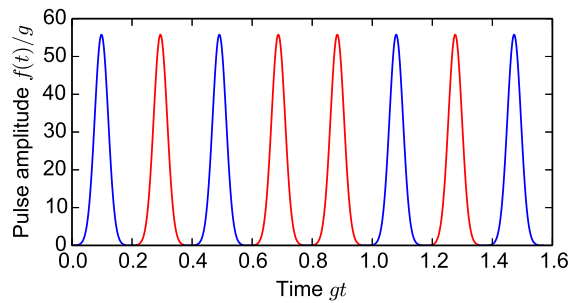}
\end{center}
\caption{\label{fig:iswap_sequence} The pulse sequence used to implement the iSWAP gate. This figure shows the pulse sequence used for both of the gate qubits, where blue signifies a pulse in $X$ direction and red signifies a pulse in $Y$ direction. Neighbouring qubits use the same pulse sequence, but with $X$ and $Y$ swapped if their coupling is to be eliminated.}
\end{figure}

We have all the necessary prerequisites to implement $\text{iSWAP}$ gates on our qubit chain. The procedure is simple: over the implementation time $T=\pi/(2g)$, we apply the sequence of eight pulses developed in section \ref{sec:combined}. Due to the selective decoupling, we can implement several iSWAP gates in parallel, provided that any two gates do not share a gate qubit. Both $\sigma_1$ and $\sigma_2$ pulses can be implemented with our pulse generator as $\pi$ pulses around the $X$ or $Y$ axis. For $\sigma_1$ pulses, the phase $\varphi_i(t)$ is chosen to be $0$, for $\sigma_2$ pulses it is chosen as $\pi/2$. The amplitude $f_i(t)$ can be any smooth function with the condition 
\begin{equation}
\int_0^{t_p} dt \, f_i(t) = \pi .
\end{equation}
The pulse implementation time $t_p$ should be made as small as possible to reduce the pulse errors. However, there are some fundamental obstacles which prevent us from making $t_p$ infinitely short. For one, a physical pulse generator will have limitations on how quickly it can steer the pulse amplitude and on the maximal achievable pulse amplitude, which in turn limits the minimal pulse duration. Additionally, the rotating frame Hamiltonian in Eq. \eqref{eq:hamiltonian} was derived in the rotating wave approximation. In order to ensure validity of this approximation, we require $1 \ll 2\omega t_p$.  Another fundamental problem is the fact
that many physical implementations of qubits are only approximately two-level systems. If we probe the physical system hard enough, which in our case means if we choose $t_p\rightarrow 0$,
eventually we will excite higher states or invoke additional interactions and thus invalidate our two-level approximation.

With that in mind, let us look at what kind of pulse duration we would have to achieve to actually implement the iSWAP gate with high fidelity. Given the implementation time $T=\pi/(2g)$ of the iSWAP gate and the necessity to implement a series of eight pulses during that time, the upper limit for the pulse time is given as $t_p \le \pi/(16g)$. In our simulation, we used pulse times 
\[t_p \in [\pi/(16g), \pi/(32g), \pi/(48g), \pi(64g), \pi/(96g)] . \] 
We simulated a qubit chain of varying length with $\Delta_i = 0$ and implemented the iSWAP gate in the middle of the chain. We used Gaussian pulse shapes for the decoupling pulses, and Fig. \ref{fig:iswap_sequence} depicts the pulse sequence used. We simulated the time-dependent Schrödinger equation for the full pulse sequence and calculated the emerging state of the qubit chain, where we then traced out all of the qubits except for the two gate qubits. The resulting state $\rho$ was then compared to the expected state $\ket{\Psi} = U_\text{iSWAP} \ket{\Psi_\text{in}}$ by means of the state fidelity \cite{Nielsen2001}
\begin{equation}
F(T) =\left\vert \bra{\Psi} \rho \ket{\Psi} \right\vert.
\end{equation}
As initial states $\ket{\Psi_\text{in}}$ we used all four basis states $\ket{00}$, $\ket{01}$, $\ket{10}$ and $\ket{11}$ and took the average over the achieved fidelities. The remaining qubits were always prepared in the state $\ket{0}$. The average fidelities depending on the pulse duration $t_p$ are given in table \ref{tab:iswap_fidelity}. The results were virtually independent of the number of total qubits $N$ in the chain. We can see that even for the longest possible pulse duration $t_p=\pi/(16g)$, the gate fidelity is quite good. 

\begin{table} 
\begin{center}
\begin{tabular}{c|c|c|c|c}
$t_p=\pi/(16g)$ & $\pi/(32g)$ & $\pi/(48g)$ & $\pi/(64g)$ & $\pi/(96g)$ \\
\hline
$0.9922$ & $0.9979$ & $0.9990$ & $0.9994$ & $0.9997$
\end{tabular}
\end{center}
\caption{Numerical simulation results for the achievable fidelity of the iSWAP gate, depending on the pulse duration $t_p$. \label{tab:iswap_fidelity}}
\end{table}

Are these pulse durations realistic? Let us consider as a concrete example two superconducting flux qubits. Flux qubits with always-on couplings of the order of $g \sim 500 \, \text{MHz}$ were realized in \cite{Coupling_two_flux_qubits},
which would allow for a fast implementation of the iSWAP gate. Additionally, flux qubits feature a rather large anharmonicity, meaning that the higher energy levels after the two qubit states are separated by a significant gap. With a typical splitting of about $5\, \text{GHz}$ between the first two levels, we could in theory have a pulse amplitude of several GHz before we risk exciting the higher states. Let us assume that we could safely employ a maximum pulse amplitude $f_\text{max} = 10 \text{GHz}$. Then the achievable minimal pulse duration for that amplitude depends on the specific pulse shape. For a Gaussian pulse like we used in our simulations we find that for $t_p=\pi/(16g)$, the required maximal pulse amplitude is $f_{\text{max}}\sim 45g$. 
However, with the assumed values of $g$ and $f_\text{max}$ for the flux qubits, we only achieve a ratio of $f_\text{max}/g \sim 20$. As a consequence, we would have to reduce the coupling constant by a factor of about $2$. Alternatively, one could also look at different pulse shapes. For example, a sine-shaped pulse would only require $f_\text{max}/g 
\sim 25$, which is much closer. However, we also found in our simulations that the sine pulse performs slightly worse in terms of achievable gate fidelity. As such, there is a compromise to be made between minimizing the gate duration $T\propto 1/g$ and maximizing the gate fidelity.

Let us assume that we choose to engineer a coupling strength of $g= 100\, \text{MHz}$, which gives us some additional reserves and allows us to aim for a pulse duration of $t_p = \pi/(32g) \approx 1 \, \text{ns}$ without exciting higher states. With the driving field frequency $\omega$ tuned to the approximate qubit level splitting of $5 \, \text{GHz}$, this pulse time is then one order of magnitude larger than $1/(2\omega)$, so that the rotating wave approximation is still valid. The implementation time of the iSWAP gate is $T \approx 16\, \text{ns}$, during which $8$ pulses need to be applied, resulting in a pulse frequency of $500\, \text{MHz}$. The requirements for our pulse generator are ambitious, but not impossible. Even more encouragingly, in recent experiments flux qubits have been demonstrated with decoherence times of the order of $10\, \mu \text{s}$ \cite{Good_Flux_Qubit_Catalani, Flux_Bylander}. This means that the gate operation time is almost three orders of magnitude faster than the decoherence time, making this procedure viable for flux qubits. Other implementations of the basic model from Sec. \ref{sec:model} may impose very different limitations.

In the interest of maximizing the fidelity, we should also point out that there exist more sophisticated pulse shapes than Gaussian or sine-shaped pulses. Some of these pulse shapes were specifically engineered to reduce their own error (see, e.g., \cite{review_nmr_pulses} for a review of NMR pulse shapes or \cite{Second_order_pulses} for a more recent design), or are less likely to excite higher states in the system \cite{Steffen2003,Motzoi2009,Pulse_Shaping_Wilhelm}. Both of these properties might help to improve the gate fidelity further. However, specifically with the self-correcting pulse shapes, the price to pay is typically a significantly higher ratio $f_\text{max}/g$ to implement a particular pulse in the same time span. Thus the qubit interaction strength $g$ would have to be reduced even further, meaning that decoherence becomes a potentially larger concern. Which pulse shape is the most adequate depends on the specific needs of a particular experiment. In our numerical simulations, Gaussian shaped pulses proved to provide a suitable compromise between achievable fidelity and required maximal pulse amplitude.

\section{Implementing high-fidelity single-qubit gates} \label{sec:sq_gates}
In addition to the two-qubit iSWAP gate, we will also need to be able to perform single-qubit gates on the individual qubits. For the implementation of the single-qubit gates, we will again make use of the pulse generator. This means that the available gate operations are given by the unitary propagator of $H_c(t)$ in \eqref{eq:hamiltonian}. In particular, we can implement rotation operations around the $X$ and $Y$ axes,
\begin{align}
R_x(\phi) &= e^{-i \sigma_1 \phi/2}, \notag \\
R_y(\phi) &= e^{-i \sigma_2 \phi/2}
\end{align}
which can be realised by choosing the phase $\varphi_i$ appropriately and engineering the pulse amplitude function such that $\int_0^{t_p} dt \, f(t) = \phi$. However, as with the decoupling pulses in section \ref{sec:dd_basics}, the gate operation is disturbed by the system Hamiltonian $H$, which limits the achievable gate fidelity. For a single-qubit gate, we typically want to achieve fidelities well above $0.99$, which is a requirement to add quantum error correction later. 

\subsection{Dynamically corrected gates with Eulerian path decoupling}

\begin{figure}
\begin{center}
\includegraphics[width=7cm]{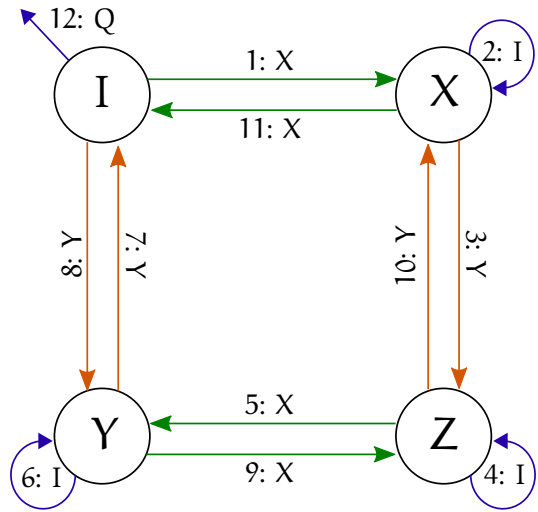}
\caption{A Eulerian path for a dynamically corrected gate operation $Q$. \label{fig:dcg}}
\end{center}
\end{figure}

Fortunately, there is a way to embed a gate operation $Q$ into a Eulerian path decoupling sequence such that it decouples the error of the gate. This technique is called dynamically corrected gates (DCG) and was introduced in \cite{khodjasteh_prl}. The idea is deceptively simple. Remember in the original Eulerian path construction in Fig. \ref{fig:euler}, each decoupling pulse formed an outgoing edge from every vertex, ensuring that its error would be decoupled to lowest order. We can add the identity operation $\id$ as another "generator" to this picture, which can be represented as loops which go out from each vertex and point back to that same node. Let us now consider that our identity operations are not perfect, but in fact given by $I = \id e^{-i\Phi_I}$, carrying an error $\Phi_I$ like the other decoupling pulses. Then this design ensures that the error is decoupled to first order. Finally, let us replace the final identity operation with the actual gate $Q$ that we want to implement, and let us 
assume that $Q$ has the same error as the faulty identity operations, $\Phi_Q = \Phi_I$. The updated graph for the resulting decoupling sequence is depicted in figure \ref{fig:dcg}. The net operation of this sequence without any errors would be the gate $Q$, as intended. Furthermore, the errors of all occurring operations are corrected to first order by the Eulerian path design.

This design hinges on the question whether we can find a faulty identity operation which has the same error as the gate $Q$. It was shown in \cite{khodjasteh_dcg} that this is possible at least to first order of the error. Consider an arbitrary gate $Q$ with its time propagator given by $U_Q(t)$ during the implementation time $t_Q$. We can introduce a scaled gate $Q_{1/2}$ with time propagator $U_{Q_{1/2}}(t) = U_Q(t/2)$, which obviously needs an implementation time of $2t_Q$ to implement the original gate $Q$. It can be shown that this scaled gate implementation carries the same error to lowest order as the faulty identity gate $I = Q^\dag Q$ with time propagator
\begin{equation}
U_I(t) = \begin{cases}
U_Q(t), & 0 \le t < t_Q, \\
U_Q(2t_q -t), & t_Q \le t \le 2t_Q.
\end{cases}
\end{equation}
In our control scheme, for any of the possible rotation gates $R_a(\varphi)$, the gates $I$ and $Q_{1/2}$ can be implemented in a straight-forward manner by modifying the phase amplitude functions $f_i(t)$. For the faulty identity gate $I$ we need 
\begin{equation}
f'_i(t) = \begin{cases}
f_i(t), & 0 \le t < t_Q, \\
-f_i(2t_Q-t), & t_Q \le t \le 2t_Q,
\end{cases}
\end{equation}
meaning that we add the negative reverse of the original pulse shape. For the gate $Q_{1/2}$ we need to scale both the time and the amplitude by $1/2$, meaning 
\begin{equation}
f'_i(t) = \frac{1}{2} f_i(t/2).
\end{equation}
If our minimal gate time is given by $t_p$, then each of the faulty $I$ operations and the final gate $Q$ will take $2t_p$ to implement. As a consequence, the total duration to implement a single-qubit gate is $16 t_p$. For the case of flux qubits as discussed in Sec. \ref{sec:iswap}, the operation times for a single gate and the iSWAP gate are comparable.

\subsection{Implementation and numerical simulations}
\begin{figure}
\begin{center}
\includegraphics[width=8cm]{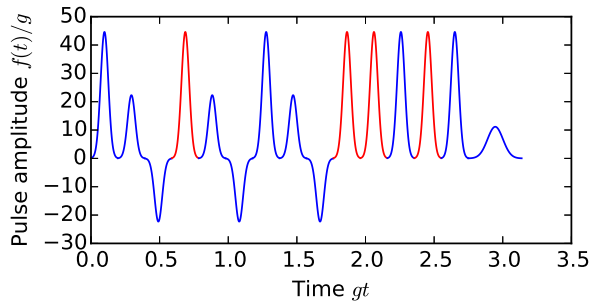}
\caption{The pulse sequence applied to a single qubit to implement a dynamically corrected $R_x(\pi/2)$ gate with Gaussian pulse shapes. Blue indicates that the pulse generator is acting along the $X$ axis, red indicates a pulse along the $Y$ axis. \label{fig:sq_pulse_sequence}}
\end{center}
\end{figure}
Figure \ref{fig:sq_pulse_sequence} shows the concrete pulse sequence we are employing in our numerical simulations to implement a dynamically corrected $R_x(\pi/2)$ gate with the decoupling sequence from figure \ref{fig:dcg}. The blue parts indicate pulses along the $X$ axis, red parts indicate pulses along the $Y$ axis. All qubits in the chain are subjected to the same sequence, except that neighbouring qubits will have the $X$ and $Y$ pulses interchanged such that the couplings between the qubits are decoupled. Qubits on which no gate is implemented will leave the pulse amplitude set to $0$ during the $I$ and $Q$ phases in the sequence. Note that several single-qubit gates can, in principle, be applied in parallel to different qubits, however not on neighbouring qubits. The reason is that on neighbouring qubits, the error associated with the gate $Q = Q_1 \otimes Q_2$ contains terms which cannot be decoupled by our decoupling scheme, and as a consequence the fidelity reduces significantly. Therefore, 
single-qubit gates on neighbouring qubits should be performed sequentially. 

As with the iSWAP gate, we simulated the pulse sequence from Fig. \ref{fig:sq_pulse_sequence} on the middle qubit of a chain with $N$ qubits by simulating the time-dependent Schrödinger equation, then tracing out all qubits but the gate qubit. The remaining traced state $\rho$ was compared to the expected state. As input states, we simulated both $\ket{0}$ and $\ket{1}$ and took the average of the resulting fidelities. The results for the implementation of the $R_x(\pi/2)$ gate can be found in table $\ref{tab:sq_fidelity}$. Results for different single-qubit gates are very similar. We can see that even for $t_p=\pi/(16g)$ the fidelity is excellent. 

\begin{table} 
\begin{center}
\begin{tabular}{c|c|c|c|c}
$t_p=\pi/(16g)$ & $\pi/(24g)$ & $\pi/(32g)$ & $\pi/(40g)$ & $\pi/(48g)$ \\
\hline
$0.99929$ & $0.99986$ & $0.99996$ & $0.99998$ & $0.99999$
\end{tabular}
\end{center}
\caption{Numerical simulation results for the achievable fidelity of the $R_x(\pi/2)$ gate, depending on the pulse duration $t_p$. \label{tab:sq_fidelity}}
\end{table}

\section{Entangling the chain qubits with the help of a CNS gate sequence} \label{sec:cnsgate}
\begin{figure} 
\begin{equation*}
\Qcircuit @C=0.8em @R=0.7em {
 & \gate{R_x(\frac{\pi}{2})} & \gate{R_y(\frac{\pi}{2})}  & \gate{R_x(-\frac{\pi}{2})}
  & \multigate{1}{\text{iSWAP}} & \gate{X} & \gate{R_y(-\frac{\pi}{2})} & \qw \\
 & \gate{R_x(\frac{\pi}{2})} & \gate{R_y(-\frac{\pi}{2})} & \qw & \ghost{\text{iSWAP}} & \qw & \qw & \qw
}
\end{equation*}
\caption{The quantum circuit to implement a CNS gate with the help of the iSWAP gate and a number of single-qubit rotations.\label{fig:cns_circuit}}
\end{figure}
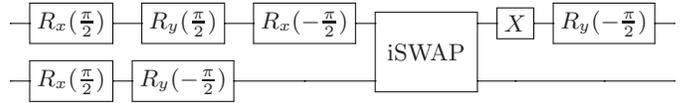

In the following, we investigate how to implement an entangling two-qubit gate in our model. An entangling gate is a necessity for universal quantum computing, and the previously implemented iSWAP gate on its own is not sufficient. However, the iSWAP gate can be combined with a sequence of single-qubit gates to perform the so called CNS gate \cite{ss03} , which is a combination of a standard CNOT followed by a SWAP operation. The gate sequence depicted in figure \ref{fig:cns_circuit} implements a $\text{CNS}$ gate with the upper qubit being the control qubit. If the control is in the state $1$, then the state of the second qubit is flipped. Afterwards, the states of both qubits are swapped. This gate is able to generate entanglement between two qubits.

In \cite{ss03}, Hadamard gates and rotations around the $Z$ axis were used. We rearranged the gate sequence to use rotations around the $X$ and $Y$ axes instead, as these are the operations accessible in our model with the help of the pulse generator.

We already have all the pieces of the puzzle to implement the CNS gate. Given that the single-qubit gates must be performed sequentially due to being on neighbouring qubits, the CNS gate will take time $\pi/(2g) + 112t_p$ to implement. For our flux qubit example with a coupling strength of $g=100\, \text{MHz}$ and $t_p = \pi/(32g)$, this yields a time of approximately $126\, \text{ns}$, which is still a factor of  80 below the decoherence time.

As a final experiment in this chapter, we will perform a sequence of CNS gates to entangle all the qubits in the chain.

\subsection{An entangling sequence of CNS gates}
If we perform a CNS gate on two qubits, of which the first (control) is prepared in the superposition $(\ket{0}+\ket{1})/\sqrt{2}$ and the second in the state $\ket{0}$, then the resulting state is $(\ket{00}+\ket{11})/\sqrt{2}$, which is an entangled Bell state. If we now take a third qubit, initially also in the state $\ket{0}$, and perform a CNS gate an qubits $2$ and $3$, then we get a three-qubit entangled state $(\ket{000}+\ket{111})/\sqrt{2}$.
With each additional execution of a CNS gate, we can bring an additional qubit into the entangled state. This type of multi-qubit entangled state is called a GHZ state \cite{ghz89}:
\begin{equation}
\ket{\text{GHZ}} = \frac{\ket{0}^{\otimes N} + \ket{1}^{\otimes N}}{\sqrt{2}}.
\end{equation}

\begin{figure} 
\begin{center}
\begin{equation*}
\Qcircuit @C=0.8em @R=0.7em {
 & \qw & \qw & \qw & \qw & \qw & \qw & \targ & \qswap & \qw \\
 & \qw & \qw & \qw & \qw & \targ & \qswap & \ctrl{-1} & \qswap \qwx & \qw \\
 & \gate{X} & \gate{R_y(-\frac{\pi}{2})} & \ctrl{1} & \qswap & \ctrl{-1} & \qswap \qwx & \qw & \qw \\
 & \qw & \qw & \targ & \qswap \qwx & \ctrl{1} & \qswap & \qw & \qw \\
 & \qw & \qw & \qw & \qw & \targ & \qswap \qwx & \ctrl{1} & \qswap & \qw \\
 & \qw & \qw & \qw & \qw & \qw & \qw & \targ & \qswap \qwx & \qw \\
}
\end{equation*}
\end{center}
\caption{A quantum circuit to entangle all qubits in a quantum register in a GHZ state. In this figure, the CNS gates are represented by a directed CNOT gate followed by a SWAP gate. \label{fig:ghz_circuit}}
\end{figure}
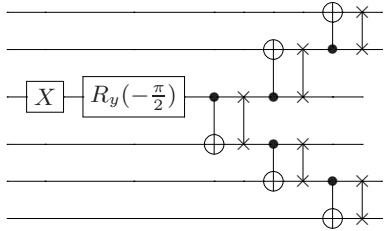

Let us assume that all qubits on the chain are initially prepared in the state $\ket{0}$. Then we bring a qubit in the middle of the chain into the superposition $(\ket{0}+\ket{1})/\sqrt{2}$. This is done by applying a Hadamard gate to it, which in our model we can express as an $X$ gate followed by a rotation $R_y\left(-\pi/2\right)$. From there on we apply CNS gates to entangle this qubit with all the other qubits in the chain, where we can in fact apply CNS gates in parallel. A gate sequence for a $6$-qubit chain is depicted in figure \ref{fig:ghz_circuit}.

We conducted numerical simulations for this gate sequence by calculating the resulting state $\ket{\Psi}$ by simulating the time-dependent Schrödinger equation, where we assume that all qubits are initially in the state $\ket{0}$. We calculated the fidelity $F_\text{GHZ}$ of the GHZ state depending on the pulse duration $t_p$ for Gaussian pulse shapes,
\begin{equation}
F_\text{GHZ} = \left\vert \braket{\text{GHZ}}{\Psi} \right\vert .
\end{equation}
We simulated qubit chains of up to 9 qubits. The results are shown in table \ref{tab:ghz_fidelity}. Given pulses which are sufficiently quick compared to the coupling strength $g$, a high fidelity of $0.99$ for the entangled state can theoretically be achieved even for $N=9$ qubits. However, at least in the flux qubit case, this would require to reduce the coupling strength $g$ to the point that the full gate sequence will approach the flux qubit decoherence time. For the more realistic pulse duration $t_p = \pi/(32g)$ the achieved fidelities are not as spectacular, but still promising.

\begin{table} 
\begin{center}
\begin{tabular}{l|r|r|r|r|r}
$N$ & $t_p=\pi/(16g)$ & $\pi/(32g)$ & $\pi/(48g)$ & $\pi/(64g)$ & $\pi/(96g)$ \\
\hline
3	& 0.964 & 0.989 & 0.995 & 0.997 & 0.999 \\
4	& 0.933 & 0.982 & 0.992 & 0.995 & 0.998 \\
5	& 0.882 & 0.974 & 0.988 & 0.993 & 0.997 \\
6	& 0.835 & 0.967 & 0.986 & 0.992 & 0.996 \\
7   & 0.821 & 0.962 & 0.983 & 0.990 & 0.996 \\
8   & 0.784 & 0.956 & 0.981 & 0.989 & 0.995 \\
9   & 0.710 & 0.947 & 0.977 & 0.987 & 0.994
\end{tabular}
\end{center}
\caption{Numerical simulation results for the achievable fidelity of the GHZ state, depending on the number $N$ of qubits and the pulse duration $t_p$. \label{tab:ghz_fidelity}}
\end{table}

It is clear that with increasing $N$, the fidelities will steadily drop. This is a consequence of the increased number of imperfect gate operations. Additionally, the longer the gate sequence, the closer we get to the decoherence time, at which point everything breaks down.
In order to achieve scalability, the addition of quantum error correction is therefore necessary. 
We believe that the demonstrated gate fidelities for single-qubit gates and the iSWAP gate are sufficiently high that error correction is feasible. For a possible implementation, we would propose to extend the qubit chain model to a two-dimensional grid, on which we could then employ a surface code. The extension to the grid requires modifications to the decoupling scheme, which are not trivial, but should be possible. Such a scenario has been accomplished recently for Ising-type qubit couplings by De and Pryadko in \cite{de_pryadko_lett, de_pryadko_toric}.

\subsection{Influence of disorder}
\begin{figure} 
\includegraphics[width=8cm]{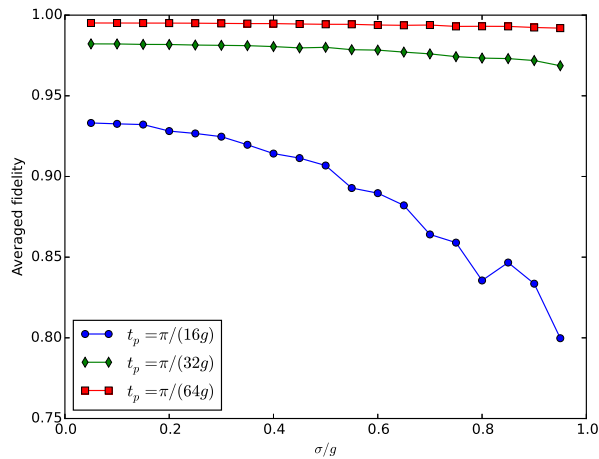}
\caption{Averaged fidelity for a GHZ state achievable on a 4-qubit chain for different values of the pulse duration $t_p$ (Gaussian pulse shapes were used), when the qubit eigenenergies differ from each other. The $\Delta_i$ are randomly sampled from a Gaussian distribution with standard deviation $\sigma$. The plotted results were averaged over 100 runs. \label{fig:disturbed}}
\end{figure}

The results in table \ref{tab:ghz_fidelity} were achieved under the assumption that the qubits' eigenergies are in resonance, meaning that the $\Delta_i$ in Eq. \eqref{eq:hamiltonian} are all zero. Non-zero $\Delta_i$ have a detrimental effect on the achievable fidelity. However, our decoupling scheme offers limited robustness against these effects. We ran additional simulations where we sampled the $\Delta_i$ randomly from a Gaussian distribution with mean value $\mu = 0$ and standard deviation $\sigma$. Results of the achievable fidelity depending on $\sigma$, averaged over 100 runs,  are plotted in Fig. \ref{fig:disturbed} for a chain of four qubits. We can see that the drop in the averaged fidelity is noticeable for $t_p=\pi/(16g)$, but with faster pulses becomes negligible, at least up to the simulated maximal value of $\sigma/g = 1$. In a recent experiment with 20 flux qubits \cite{Metamaterial}, deviations of up to $1\, \text{GHz}$ were observed in the eigenenergies, which may be two to ten times larger than the coupling $g$, depending on how strongly the qubits are engineered to interact. As such, current experimental deviations may be larger than our decoupling scheme can handle. However, we expect that with improved manufacturing processes the qubit eigenenergy discrepancies will become sufficiently small in the future so that the detrimental influence of the disorder is negligible with sufficiently fast pulses.

\section{Conclusions}
We presented a coupled qubit system modelled after superconducting flux qubits which is fully controlled by a pulse generator. The qubits are strongly coupled to their neighbours, and the coupling is always present. We demonstrated how the pulse generator can be used to implement both single-qubit rotations and the two-qubit iSWAP gate. For the implementation of the two-qubit gate we exploit the coupling between the qubits and use a Eulerian decoupling scheme to decouple the gate qubits from the remaining qubits in the system. The decoupling scheme is flexible so that several two-qubit gates can be implemented in parallel. The single-qubit rotations are realised with the help of dynamically corrected gate operations, which embed the gate operation into a Eulerian decoupling sequence.

The efficiency of our control scheme was analyzed in numerical simulations, where we first looked at single gate applications and achieved high fidelities for both the iSWAP gate and the single-qubit rotations. Then a sequence of CNS gates was simulated to entangle all the qubits in the chain in a GHZ state. In order to entangle $N$ qubits in a GHZ state, $N-1$ CNS gates are required. Without error correction, the GHZ state fidelity directly depends on the number of qubits in the chain. We found that for sufficiently short pulses, we could still achieve a fidelity of $0.99$ and above for chains of up to $N=9$ qubits. However, the pulse length is physically limited by the energy gap to higher excited states, which should not be excited by the pulse generator. As a consequence, the coupling strengths between the qubits may need to be reduced, which in turn increases the gate implementation times and could cause problems with decoherence.

In order to achieve true scalability, error correction will be required. De and Pryadko recently demonstrated how a universal set of quantum gates could be implemented on a qubit lattice with Ising couplings and then implemented the toric code on top of this lattice to achieve scalability \cite{de_pryadko_toric}. We believe that this approach could be adopted in principle for our model.

\acknowledgments
The authors acknowledge financial support by CASEDIII and by the BMBF-project Q.com. This work has been co-funded by the DFG as part of project P4 within the CRC 1119 CROSSING.

\end{document}